\documentclass[preprint,aps,showpacs,showkeys,amsmath,amssymb,nopreprintnumbers,nofootinbib]{revtex4}
\usepackage{graphicx}
\usepackage{dcolumn}
\usepackage{bm}
\usepackage[dvipdfm]{color}
\usepackage[dvipdfm,bookmarks=true,bookmarksnumbered=true,bookmarkstype=toc]{hyperref}

\def\Rnum#1{\resizebox{0.5em}{\height}{\uppercase\expandafter{\romannumeral #1}}}

\begin{document}


\title{Neutron stars with Hyperons \\
in Dirac-Brueckner-Hartree-Fock approach}%

\author{Tetsuya Katayama}
\email{6213701@ed.tus.ac.jp}
\affiliation{%
Department of Physics, Faculty of Science and Technology,\\
Tokyo University of Science, Noda 278-8510, Japan 
}%
\author{Koichi Saito}
\email{koichi.saito@rs.tus.ac.jp}
\altaffiliation[Also at ]{J-PARC Branch, KEK Theory Center, Institute of Particle and Nuclear Studies, KEK, Tokai 319-1106, Japan}
\affiliation{%
Department of Physics, Faculty of Science and Technology,\\
Tokyo University of Science, Noda 278-8510, Japan 
}%

\date{\today}

\begin{abstract}
Using the Dirac-Brueckner-Hartree-Fock (DBHF) approach including the hyperon degrees of freedom, we investigate the properties of neutron-star matter.  To handle the hyperons in matter, we first examine the importance of the space part of baryon self-energies at high densities, and secondly study the effect of negative-energy states of baryons, which can provide an unambiguous relationship between the in-medium reaction matrices for baryon-baryon scattering and the baryon self-energies.  We solve the coupled, Bethe-Salpeter equations in the nuclear-matter rest frame by using the Bonn potentials.  We assume that eight kinds of nonstrange and strange mesons ($\sigma,\,\delta,\,\omega,\,\rho,\,\eta,\,\pi,\,K,\,K^{\ast}$) take part in the interactions between two baryons. 
Then, we calculate the baryon self-energies, the energy density and pressure of matter.  The present calculation provides a hard equation of state in neutron-star matter at high densities, which is generated by the effect of Pauli exclusion, the short-range correlations between two baryons, etc.  We finally predict the maximum neutron-star mass of $2.02\,M_{\odot}$, which is consistent with both the recently observed masses, $1.97\pm0.04M_{\odot}$ (J1614-2230) and $2.01\pm0.04M_{\odot}$ (J0348+0432).  
\end{abstract}


\maketitle
\section{Introduction \label{sec:intro}}
Neutron stars may be the most dense and exotic state of nuclear matter in the universe.  In particular, the core of 
neutron star serves as a natural laboratory to study the extreme situation of nuclear matter whose density reaches several times higher than the normal nuclear density, $n_B^0$ \cite{Glendenning}.
The recent observations of the massive neutron stars, J1614-2230 with the mass of $1.97\pm0.04M_{\odot}$  ($M_{\odot}$: the solar mass) \cite{J16142230} and J0348+0432 ($2.01\pm0.04M_{\odot}$) \cite{J03480432}, have provided the tight constraints on the equation of state (EoS) for dense nuclear matter.  

To understand these massive objects, 
various nuclear models have been examined, in which relativistic mean-field theory (RMFT) or relativistic Hartree (RH) approximation \cite{MFT} is very popular and has been successfully applied to describe not only the properties of nuclear matter but also finite nuclei. 
However, in the RH calculation for the EoS of neutron stars, it is necessary to extrapolate the EoS around $n_B^0$ to 
the high-density region, where the short-range correlations among baryons become very important.

To remedy this defect, one needs to proceed to the Dirac-Brueckner-Hartree-Fock (DBHF) approach.\footnote{
Of course, the short-range correlations can be treated within non-relativistic formulation as well \cite{BHF}.  
However, in matter 
at much higher densities than $n_B^0$, relativity becomes inevitable. }
In the DBHF calculations \cite{DBHF1,DBHF2,DBHF3,Fuchs1,Fuchs2,Fuchs3,Dalen,without_PW,Poschenrieder,Huber,Jong}, the properties of nuclear matter around $n_B^0$ can be described well by using several nucleon-meson coupling constants 
determined by an enormous amount of the elastic and inelastic nucleon-nucleon scattering data. 
Thus, in contrast to RMFT, there is, in principle, no adjustable parameters to fit the matter properties. 
In the DBHF approach, it is well recognized that the density dependence of in-medium two-nucleon scattering amplitudes 
is mainly generated by the Pauli exclusion principle and the short-range correlations, and that such density dependence 
plays a key role in nuclear matter.  Furthermore, the DBHF method effectively takes into account 
a particular class of three-body forces, and thus it can explain the nuclear saturation in symmetric nuclear matter \cite{Fuchs3}. 
The DBHF approach may be the most appropriate method to calculate the EoS for dense nuclear matter like neutron stars, 
and it is quite interesting to perform such calculations including the degrees of freedom of hyperons. 

Until now, several groups have carried out the DBHF calculations not only in the region around $n_B^0$ but also in matter at higher densities  \cite{DBHF1,DBHF2,DBHF3,Fuchs1,Fuchs2,Fuchs3,Dalen,without_PW}.  
However, if we want to take hyperons into account as well as nucleon, the conventional 
approach is not convenient, because it usually requires the following, two assumptions: (1) the space component of 
the in-medium baryon self-energy, $\Sigma^V$, is supposed to be small and hence ignored,\footnote{
In Refs.\cite{DBHF2,DBHF3,Fuchs1,Fuchs2,Fuchs3,Dalen,Jong}, the $\Sigma^V$ is partly taken into account in the in-medium nucleon mass, which is called the reduced effective mass.} (2) the relationship between the in-medium on-shell T-matrix for nucleon-nucleon scattering and the in-medium nucleon self-energies is not clear \cite{Fuchs1,Fuchs2,Fuchs3}. 

Concerning the assumption (1), up to around $n_B^0$, we can certainly expect that the space part of the self-energy is small compared with the other components (see, for example, Ref.\cite{Fuchs2}).  However, we do not have any warranty on this assumption in extremely dense matter.  
As for the assumption (2), the ambiguity emerges from limiting the state vectors of nucleon in matter, that is, only the positive-energy states are usually considered, while the negative-energy ones are discarded.  Thus, if we can lift this restriction, the problem may be settled. 
In fact, in Refs.\cite{Poschenrieder,Huber,Jong}, the negative-energy states are partly included in the Bethe-Salpeter equation, 
and this method seems to be successful in removing the ambiguity.  

Furthermore, in the usual DBHF calculation, it is often assumed that 
the mass difference between the interacting proton and neutron can be ignored even in isospin-asymmetric nuclear matter \cite{DBHF3,Dalen}.\footnote{
Strictly speaking, the mass difference is neglected in the Born term of the Bethe-Salpeter equation.}  However, in general, the difference between those masses is likely to be large and may not be negligible.  The proton-neutron mass difference in matter has been already considered in Ref.\cite{Huber}.  
When hyperons take place in dense nuclear matter, the effective masses of interacting two baryons are markedly different from each other.  

Based on the above considerations, in the present paper, we solve the coupled, Bethe-Salpeter equations in the nuclear-matter rest frame.    
We here adopt the Bonn potentials \cite{DBHF1}.  Then, we calculate the baryon self-energies, the energy density and pressure of matter.  We also study the momentum dependence of the nucleon self-energies in details.
Including the hyperon degrees of freedom, we finally estimate the possible, maximum mass of neutron stars, which may be the first ever result in the DBHF calculation. 

The paper is organized as follows. In Sec. \ref{sec:formalism}, we briefly describe the present method of the DBHF calculation.
In Sec. \ref{sec:symmetric_nuclear_matter}, we compare the present DBHF method to the conventional one, and discuss the problems mentioned above. 
Secs. \ref{protoneutron_stars} and \ref{including_hyperons} deal with the equations of state for neutron stars without and with hyperons, respectively.
The conclusions are presented in Sec. \ref{sec:summary_and_conclusion}.

\section{Method \label{sec:formalism}}
In order to describe the baryon-baryon scattering in nuclear matter, we start with the self-energy of baryon ($B$) in the rest frame of infinite, uniform nuclear matter.  It is given by 
\begin{equation}
    \Sigma_B(k)=\Sigma_B^{S}(k)-\gamma_0\Sigma_B^0(k)+{\boldsymbol\gamma}\cdot\bm{k}\Sigma_B^V(k),  \label{eq_S}
\end{equation}
where $\bm{k} \, (k)$ is the three- (four-) momentum of baryon.  Here, $\Sigma_B^{S \, (0) \, [V]}$ is the scalar (the zero-th component of vector) [the space component of vector] part of the baryon self-energy.
Using these self-energies, the effective mass, $M_B^{\ast}$, the effective momentum, $\bm{k}_B^{\ast}$, and the effective energy, $E^{\ast}_B$, of baryon in matter are defined as
\begin{eqnarray}
    M_B^{\ast}(k)&\equiv&M_B+\Sigma_B^S(k),\label{eq_MsB}\\
    \bm{k}^{\ast}_B&\equiv&\bm{k}[1+\Sigma_B^V(k)],\\
    E^{\ast}_B(k)&\equiv&\sqrt{\bm{k}_B^{\ast2}+M_B^{\ast2}(k)},\label{eq_EsB}
\end{eqnarray}
with $M_B$ being the free baryon mass.
It should be noted that, because in this work we fully consider the space component of the self-energy, $\Sigma_B^V$, the definitions are different from those in Refs.\cite{DBHF1,DBHF2,DBHF3,Fuchs1,Fuchs2,Fuchs3,Dalen,without_PW,Jong}. 

Then, the baryon spinor states with positive or negative energy in matter are respectively constructed as 
\begin{eqnarray}
    \Phi_B(\bm{k},s)&=&\sqrt{M_B^{\ast}(k)+E^{\ast}_B(k)}\left(
    \begin{array}{c}
        \chi_s ,\\
        \frac{\bm{k}_B^{\ast}\cdot{\boldsymbol\sigma}}{M_B^{\ast}(k)+E_B^{\ast}(k)}\chi_{s}
    \end{array}
    \right)\label{eq_positive_spinor},\\
    \Theta_B(\bm{k},s)&=&\sqrt{M_B^{\ast}(k)+E^{\ast}_B(k)}\left(
    \begin{array}{c}
        \frac{\bm{k}_B^{\ast}\cdot{\boldsymbol\sigma}}{M_B^{\ast}(k)+E_B^{\ast}(k)}\chi_{-s} \\
        \chi_{-s}
    \end{array}
    \right)\label{eq_negative_spinor},
\end{eqnarray}
where ${\boldsymbol \sigma}$ is the Pauli matrix, and $\chi_s$ denotes a 2-component Pauli spinor. 

In the conventional DBHF method, the baryon-baryon scattering is usually evaluated in the center of mass frame with respect to the interacting two baryons. 
In such a case, instead of the standard spinors, Eqs.(\ref{eq_positive_spinor})-(\ref{eq_negative_spinor}), the helicity spinors and the  partial-wave decomposition are often used to calculate the Bethe-Salpeter equation \cite{DBHF1,DBHF2,DBHF3,Fuchs1,Fuchs2,Fuchs3,Dalen,Poschenrieder,Huber,Jong}.  
This method is certainly convenient when the space part of the self-energy, $\Sigma_B^V$, can be insignificant. 
However, in the case where $\Sigma_B^V$ remains and $\bm{k}^{\ast}_B \neq \bm{k}$, we may face a new trouble. 

\begin{figure}[htbp]
 \begin{center}
  \includegraphics[width=155mm]{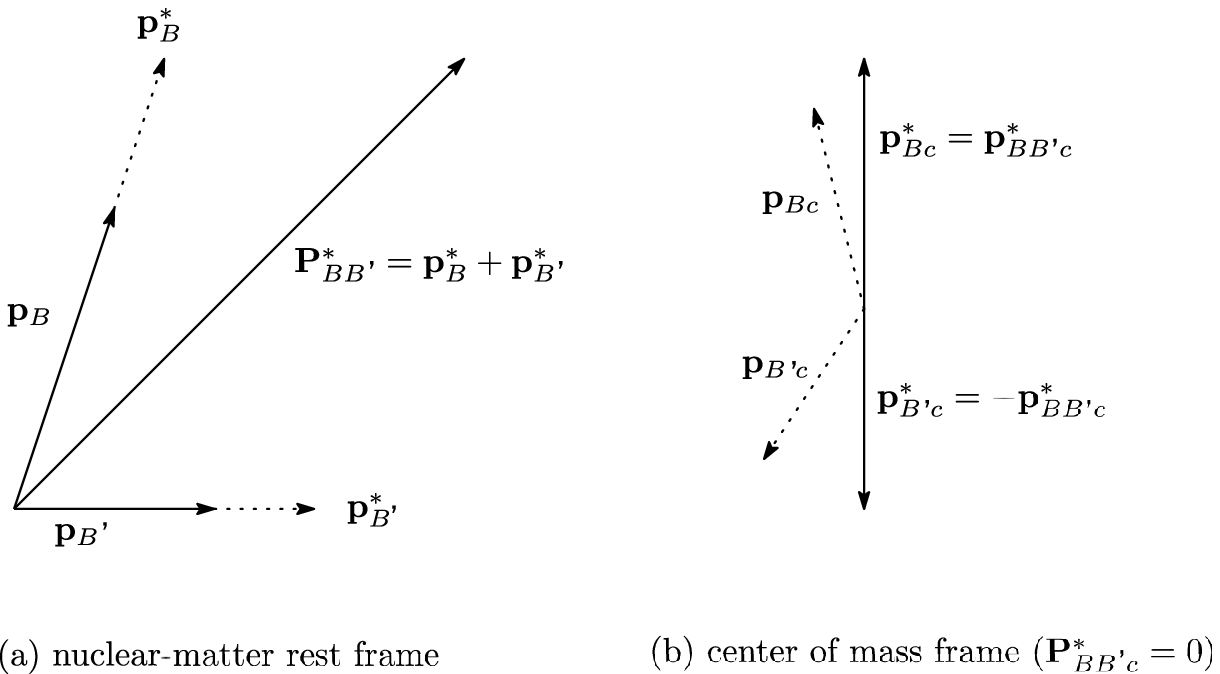}
 \end{center}
 \caption{Momenta in the nuclear-matter rest frame (a), and in the center of mass frame (b).}
 \label{fig_Schematic_representation_of_momenta}
\end{figure}

To illustrate it, let us assume that the (original) three-momenta and the effective ones of the interacting two particles, $B$ and $B^{\prime}$, are respectively given by ($\bm{p}_B, \, \bm{p}_B^{\ast}$) and ($\bm{p}_{B^{\prime}}, \, \bm{p}_{B^{\prime}}^{\ast}$)  in the nuclear-matter rest frame, where $\bm{p}_{B(B^{\prime})}$ and $\bm{p}_{B(B^{\prime})}^{\ast}$ are parallel each other 
(see Fig.\ref{fig_Schematic_representation_of_momenta} (a)).  Performing the Lorentz boost which is along the effective total three-momentum, $\bm{P}_{BB^{\prime}}^{\ast}=\bm{p}_B^{\ast}+\bm{p}_{B^{\prime}}^{\ast}$, and is chosen so that the two effective three-momenta of the interacting particles, $\bm{p}_{Bc}^{\ast}$ and $\bm{p}_{B^{\prime}c}^{\ast}$, are equal and opposite in the center of mass frame, namely $\bm{P}_{BB^{\prime}c}^{\ast}=0$ (the momenta in the center of mass frame are denoted with subscript $c$), the system is transferred from the matter-rest frame to the center of mass frame (see Fig.\ref{fig_Schematic_representation_of_momenta} (b)).  
Then, as seen in Fig.\ref{fig_Schematic_representation_of_momenta}, 
the direction of three-momentum, $\bm{p}_{Bc (B^{\prime}c)}$, does not in general coincide with the effective one, $\bm{p}^{\ast}_{Bc (B^{\prime}c)}$.
Thus, even if the helicity spinor, which satisfies ${\boldsymbol\sigma}\cdot\bm{p}_{B(B^{\prime})c}^{\ast}/|\bm{p}_{B(B^{\prime})c}^{\ast}|\left|l\right>=2l\left|l\right>$ with eigenvalues $l=\pm\frac{1}{2}$, is introduced, it cannot fulfill the condition, ${\boldsymbol\sigma}\cdot\bm{p}_{B(B^{\prime})c}/|\bm{p}_{B(B^{\prime})c}|\left|l\right>=2l\left|l\right>$, simultaneously. 

Because the self energy, $\Sigma_B^V$, may become significant at high densities and we want to fully incorporate it into our calculation, it is thus more reasonable to carry out the calculation with the standard spinors, Eqs.(\ref{eq_positive_spinor})-(\ref{eq_negative_spinor}), in the nuclear-matter rest frame, rather than with the helicity spinors in the center of mass frame.  

It is also worth noting that the inclusion of the negative-energy states of baryon in the Bethe-Salpeter amplitude may be vital to remove the ambiguity of the relationship between the in-medium reaction matrices for baryon-baryon scattering and the in-medium baryon self-energies \cite{Poschenrieder,Huber,Jong}.  As discussed in the previous section, such ambiguity may arise from the restriction that only the positive-energy states are taken into account in the calculation \cite{DBHF1,DBHF2,DBHF3,Fuchs1,Fuchs2,Fuchs3,Dalen,without_PW}. 
If one includes the negative-energy states as well as the positive ones, the relationship can be uniquely determined \cite{Poschenrieder,Huber,Jong}.  

\begin{figure}[htbp]
 \begin{center}
  \includegraphics[width=150mm]{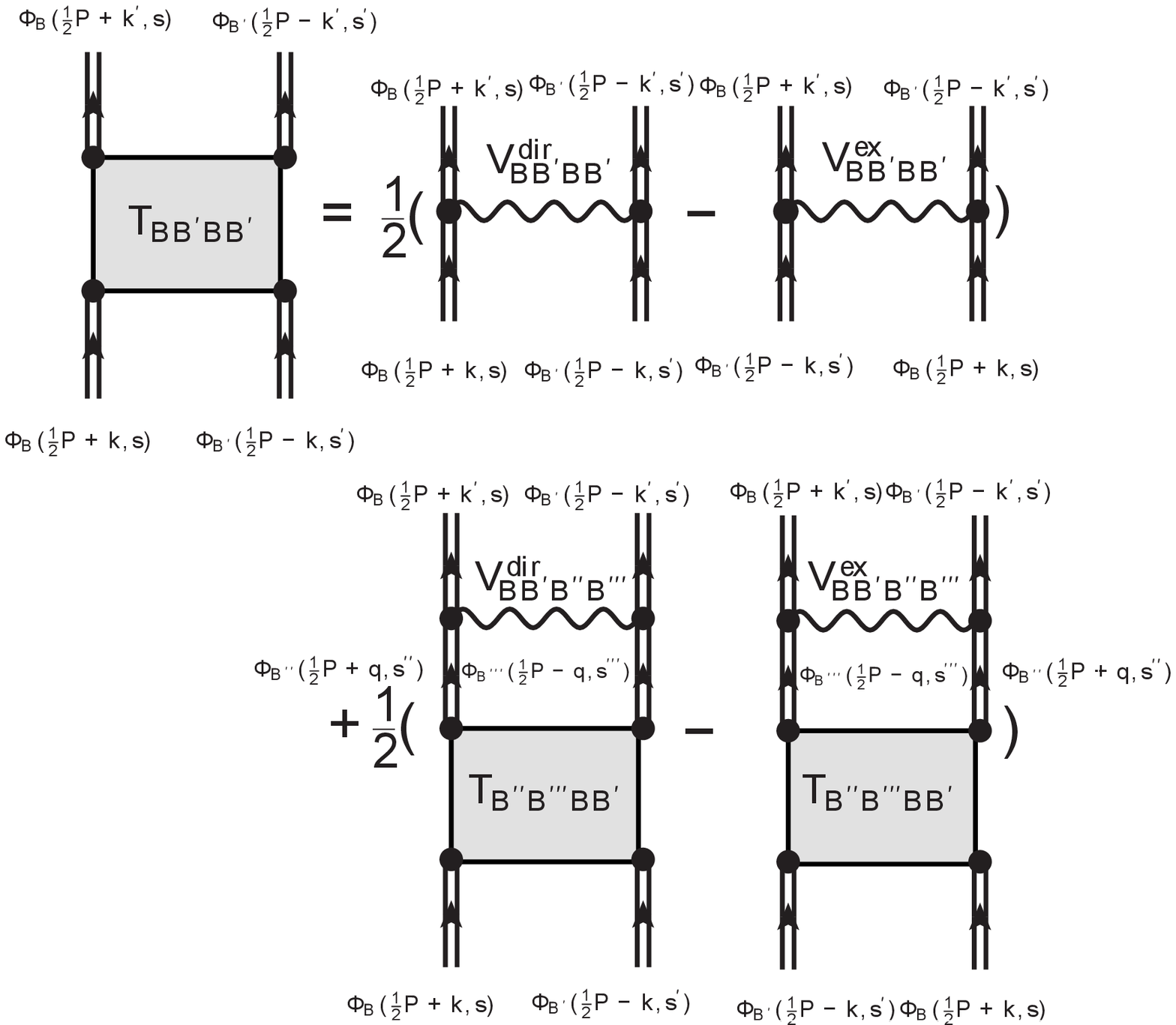}
 \end{center}
 \caption{As an example, the Feynman diagram of reaction amplitude, $T$, is shown explicitly.}
 \label{fig_BS_eq_for_T}
\end{figure}

We thus define four reaction amplitudes 
\begin{eqnarray}
    &&T_{B^{\prime\prime\prime}B^{\prime\prime}B^{\prime}B}(\bm{k}^{\prime},\bm{k},s^{\prime\prime\prime},s^{\prime\prime},s^{\prime},s;\bm{P})\nonumber\\
    &\equiv&\bar{\Phi}_{B^{\prime\prime\prime}}\left(\frac{1}{2}\bm{P}+\bm{k}^{\prime},s^{\prime\prime\prime}\right)\bar{\Phi}_{B^{\prime\prime}}\left(\frac{1}{2}\bm{P}-\bm{k}^{\prime},s^{\prime\prime}\right)\Gamma\Phi_{B^{\prime}}\left(\frac{1}{2}\bm{P}+\bm{k},s^{\prime}\right)\Phi_B\left(\frac{1}{2}\bm{P}-\bm{k},s\right),\\
    &&R_{B^{\prime\prime\prime}B^{\prime\prime}B^{\prime}B}(\bm{k}^{\prime},\bm{k},s^{\prime\prime\prime},s^{\prime\prime},s^{\prime},s;\bm{P})\nonumber\\
    &\equiv&\bar{\Theta}_{B^{\prime\prime\prime}}\left(\frac{1}{2}\bm{P}+\bm{k}^{\prime},s^{\prime\prime\prime}\right)\bar{\Phi}_{B^{\prime\prime}}\left(\frac{1}{2}\bm{P}-\bm{k}^{\prime},s^{\prime\prime}\right)\Gamma\Phi_{B^{\prime}}\left(\frac{1}{2}\bm{P}+\bm{k},s^{\prime}\right)\Phi_B\left(\frac{1}{2}\bm{P}-\bm{k},s\right),\\
    &&O_{B^{\prime\prime\prime}B^{\prime\prime}B^{\prime}B}(\bm{k}^{\prime},\bm{k},s^{\prime\prime\prime},s^{\prime\prime},s^{\prime},s;\bm{P})\nonumber\\
    &\equiv&\bar{\Phi}_{B^{\prime\prime\prime}}\left(\frac{1}{2}\bm{P}+\bm{k}^{\prime},s^{\prime\prime\prime}\right)\bar{\Phi}_{B^{\prime\prime}}\left(\frac{1}{2}\bm{P}-\bm{k}^{\prime},s^{\prime\prime}\right)\Gamma\Theta_{B^{\prime}}\left(\frac{1}{2}\bm{P}+\bm{k},s^{\prime}\right)\Phi_B\left(\frac{1}{2}\bm{P}-\bm{k},s\right),\\
    &&P_{B^{\prime\prime\prime}B^{\prime\prime}B^{\prime}B}(\bm{k}^{\prime},\bm{k},s^{\prime\prime\prime},s^{\prime\prime},s^{\prime},s;\bm{P})\nonumber\\
    &\equiv&\bar{\Theta}_{B^{\prime\prime\prime}}\left(\frac{1}{2}\bm{P}+\bm{k}^{\prime},s^{\prime\prime\prime}\right)\bar{\Phi}_{B^{\prime\prime}}\left(\frac{1}{2}\bm{P}-\bm{k}^{\prime},s^{\prime\prime}\right)\Gamma\Theta_{B^{\prime}}\left(\frac{1}{2}\bm{P}+\bm{k},s^{\prime}\right)\Phi_B\left(\frac{1}{2}\bm{P}-\bm{k},s\right),
\end{eqnarray}
where $\Gamma$ represents the effective reaction operator, and these amplitudes satisfy the following, coupled Bethe-Salpeter equations
\begin{eqnarray}
    &&T_{BB^{\prime}BB^{\prime}}(\bm{k},\bm{k},s,s^{\prime},s,s^{\prime};\bm{P})=\bar{V}_{BB^{\prime}BB^{\prime}}(\bm{k},\bm{k},s,s^{\prime},s,s^{\prime};\bm{P})\nonumber\\
    &+&\sum_{s^{\prime\prime}s^{\prime\prime\prime}B^{\prime\prime}B^{\prime\prime\prime}}\int\frac{d^3q}{(2\pi)^4}\bar{V}_{BB^{\prime}B^{\prime\prime}B^{\prime\prime\prime}}(\bm{k},\bm{q},s,s^{\prime},s^{\prime\prime},s^{\prime\prime\prime};\bm{P})\nonumber\\
    &&\times Q_{B^{\prime\prime}B^{\prime\prime\prime}}(\bm{P},\bm{q})g_{Th\,B^{\prime\prime}B^{\prime\prime\prime}}(\bm{P},\bm{q})T_{B^{\prime\prime}B^{\prime\prime\prime}BB^{\prime}}(\bm{q},\bm{k},s^{\prime\prime\prime},s^{\prime\prime},s,s^{\prime};\bm{P}),\label{eq_BS_for_T}\\
    &&R_{BB^{\prime}BB^{\prime}}(\bm{k},\bm{k},s,s^{\prime},s,s^{\prime};\bm{P})=\bar{U}_{BB^{\prime}BB^{\prime}}(\bm{k},\bm{k},s,s^{\prime},s,s^{\prime};\bm{P})\nonumber\\
    &+&\sum_{s^{\prime\prime}s^{\prime\prime\prime}B^{\prime\prime}B^{\prime\prime\prime}}\int\frac{d^3q}{(2\pi)^4}\bar{U}_{BB^{\prime}B^{\prime\prime}B^{\prime\prime\prime}}(\bm{k},\bm{q},s,s^{\prime},s^{\prime\prime},s^{\prime\prime\prime};\bm{P})\nonumber\\
    &&\times Q_{B^{\prime\prime}B^{\prime\prime\prime}}(\bm{P},\bm{q})g_{Th\,B^{\prime\prime}B^{\prime\prime\prime}}(\bm{P},\bm{q})T_{B^{\prime\prime}B^{\prime\prime\prime}BB^{\prime}}(\bm{q},\bm{k},s^{\prime\prime\prime},s^{\prime\prime},s,s^{\prime};\bm{P}),\label{eq_BS_for_R}\\
    &&O_{BB^{\prime}BB^{\prime}}(\bm{k},\bm{k},s,s^{\prime},s,s^{\prime};\bm{P})=\bar{W}_{BB^{\prime}BB^{\prime}}(\bm{k},\bm{k},s,s^{\prime},s,s^{\prime};\bm{P})\nonumber\\
    &+&\sum_{s^{\prime\prime}s^{\prime\prime\prime}B^{\prime\prime}B^{\prime\prime\prime}}\int\frac{d^3q}{(2\pi)^4}\bar{V}_{BB^{\prime}B^{\prime\prime}B^{\prime\prime\prime}}(\bm{k},\bm{q},s,s^{\prime},s^{\prime\prime},s^{\prime\prime\prime};\bm{P})\nonumber\\
    &&\times Q_{B^{\prime\prime}B^{\prime\prime\prime}}(\bm{P},\bm{q})g_{Th\,B^{\prime\prime}B^{\prime\prime\prime}}(\bm{P},\bm{q})O_{B^{\prime\prime}B^{\prime\prime\prime}BB^{\prime}}(\bm{q},\bm{k},s^{\prime\prime\prime},s^{\prime\prime},s,s^{\prime};\bm{P}),\label{eq_BS_for_O}\\
    &&P_{BB^{\prime}BB^{\prime}}(\bm{k},\bm{k},s,s^{\prime},s,s^{\prime};\bm{P})=\bar{Z}_{BB^{\prime}BB^{\prime}}(\bm{k},\bm{k},s,s^{\prime},s,s^{\prime};\bm{P})\nonumber\\
    &+&\sum_{s^{\prime\prime}s^{\prime\prime\prime}B^{\prime\prime}B^{\prime\prime\prime}}\int\frac{d^3q}{(2\pi)^4}\bar{U}_{BB^{\prime}B^{\prime\prime}B^{\prime\prime\prime}}(\bm{k},\bm{q},s,s^{\prime},s^{\prime\prime},s^{\prime\prime\prime};\bm{P})\nonumber\\
    &&\times Q_{B^{\prime\prime}B^{\prime\prime\prime}}(\bm{P},\bm{q})g_{Th\,B^{\prime\prime}B^{\prime\prime\prime}}(\bm{P},\bm{q})O_{B^{\prime\prime}B^{\prime\prime\prime}BB^{\prime}}(\bm{q},\bm{k},s^{\prime\prime\prime},s^{\prime\prime},s,s^{\prime};\bm{P}),\label{eq_BS_for_P}
\end{eqnarray}
with
\begin{eqnarray}
    &&\bar{V}_{B^{\prime\prime\prime}B^{\prime\prime}B^{\prime}B}(\bm{k}^{\prime},\bm{k},s^{\prime\prime\prime},s^{\prime\prime},s^{\prime},s;\bm{P})\nonumber\\
    &\equiv&\frac{1}{2}\left[V^{dir}_{B^{\prime\prime\prime}B^{\prime\prime}B^{\prime}B}(\bm{k}^{\prime},\bm{k},s^{\prime\prime\prime},s^{\prime\prime},s^{\prime},s;\bm{P})-V^{ex}_{B^{\prime\prime\prime}B^{\prime\prime}B^{\prime}B}(\bm{k}^{\prime},\bm{k},s^{\prime\prime\prime},s^{\prime\prime},s^{\prime},s;\bm{P})\right],\\
    &&\bar{U}_{B^{\prime\prime\prime}B^{\prime\prime}B^{\prime}B}(\bm{k}^{\prime},\bm{k},s^{\prime\prime\prime},s^{\prime\prime},s^{\prime},s;\bm{P})\nonumber\\
    &\equiv&\frac{1}{2}\left[U^{dir}_{B^{\prime\prime\prime}B^{\prime\prime}B^{\prime}B}(\bm{k}^{\prime},\bm{k},s^{\prime\prime\prime},s^{\prime\prime},s^{\prime},s;\bm{P})-U^{ex}_{B^{\prime\prime\prime}B^{\prime\prime}B^{\prime}B}(\bm{k}^{\prime},\bm{k},s^{\prime\prime\prime},s^{\prime\prime},s^{\prime},s;\bm{P})\right],\\
    &&\bar{W}_{B^{\prime\prime\prime}B^{\prime\prime}B^{\prime}B}(\bm{k}^{\prime},\bm{k},s^{\prime\prime\prime},s^{\prime\prime},s^{\prime},s;\bm{P})\nonumber\\
    &\equiv&\frac{1}{2}\left[W^{dir}_{B^{\prime\prime\prime}B^{\prime\prime}B^{\prime}B}(\bm{k}^{\prime},\bm{k},s^{\prime\prime\prime},s^{\prime\prime},s^{\prime},s;\bm{P})-W^{ex}_{B^{\prime\prime\prime}B^{\prime\prime}B^{\prime}B}(\bm{k}^{\prime},\bm{k},s^{\prime\prime\prime},s^{\prime\prime},s^{\prime},s;\bm{P})\right],\\
    &&\bar{Z}_{B^{\prime\prime\prime}B^{\prime\prime}B^{\prime}B}(\bm{k}^{\prime},\bm{k},s^{\prime\prime\prime},s^{\prime\prime},s^{\prime},s;\bm{P})\nonumber\\
    &\equiv&\frac{1}{2}\left[Z^{dir}_{B^{\prime\prime\prime}B^{\prime\prime}B^{\prime}B}(\bm{k}^{\prime},\bm{k},s^{\prime\prime\prime},s^{\prime\prime},s^{\prime},s;\bm{P})-Z^{ex}_{B^{\prime\prime\prime}B^{\prime\prime}B^{\prime}B}(\bm{k}^{\prime},\bm{k},s^{\prime\prime\prime},s^{\prime\prime},s^{\prime},s;\bm{P})\right], 
\end{eqnarray}
where these are the anti-symmetrized one-boson-exchange (OBE) interaction kernels, as given in Ref.\cite{Gross}. 
Here, $Q_{BB^{\prime}}$ is the Pauli exclusion operator for baryons $B$ and $B^\prime$, and $g_{ThBB^{\prime}}$ denotes the  Thompson's two-particle propagator \cite{Thompson_propagator}. 
The seven arguments in the reaction amplitudes, $T_{BB^{\prime}BB^{\prime}},\,R_{BB^{\prime}BB^{\prime}},\,O_{BB^{\prime}BB^{\prime}}$ and $P_{BB^{\prime}BB^{\prime}}$, and the OBE interactions are as follows (see also Fig.\ref{fig_BS_eq_for_T}): 
from left to right, the first variable represents the final (or intermediate) relative three-momentum; the second, the initial (or intermediate) relative three-momentum; the third and fourth are for the spins of the final (or intermediate) two baryons, each of which is up ($+$) or down ($-$); the fifth and sixth, the spins of the initial (or intermediate) two baryons; and the last one is the total three-momentum of the system.
Then, the anti-symmetrized OBE interaction kernels are explicitly defined by
\begin{eqnarray}
    &&V^{dir}_{B^{\prime\prime\prime}B^{\prime\prime}B^{\prime}B}(\bm{k}^{\prime},\bm{k},s^{\prime\prime\prime},s^{\prime\prime},s^{\prime},s;\bm{P})\nonumber\\
    &\equiv&\bar{\Phi}_{B^{\prime\prime\prime}}\left(\frac{1}{2}\bm{P}+\bm{k}^{\prime},s^{\prime\prime\prime}\right)\bar{\Phi}_{B^{\prime\prime}}\left(\frac{1}{2}\bm{P}-\bm{k}^{\prime},s^{\prime\prime}\right)V\Phi_{B^{\prime}}\left(\frac{1}{2}\bm{P}+\bm{k},s^{\prime}\right)\Phi_B\left(\frac{1}{2}\bm{P}-\bm{k},s\right),\\
    &&V^{ex}_{B^{\prime\prime\prime}B^{\prime\prime}B^{\prime}B}(\bm{k}^{\prime},\bm{k},s^{\prime\prime\prime},s^{\prime\prime},s^{\prime},s;\bm{P})\nonumber\\
    &\equiv&\bar{\Phi}_{B^{\prime\prime\prime}}\left(\frac{1}{2}\bm{P}+\bm{k}^{\prime},s^{\prime\prime\prime}\right)\bar{\Phi}_{B^{\prime\prime}}\left(\frac{1}{2}\bm{P}-\bm{k}^{\prime},s^{\prime\prime}\right)V\Phi_B\left(\frac{1}{2}\bm{P}-\bm{k},s\right)\Phi_{B^{\prime}}\left(\frac{1}{2}\bm{P}+\bm{k},s^{\prime}\right),\\
    &&U^{dir}_{B^{\prime\prime\prime}B^{\prime\prime}B^{\prime}B}(\bm{k}^{\prime},\bm{k},s^{\prime\prime\prime},s^{\prime\prime},s^{\prime},s;\bm{P})\nonumber\\
    &\equiv&\bar{\Theta}_{B^{\prime\prime\prime}}\left(\frac{1}{2}\bm{P}+\bm{k}^{\prime},s^{\prime\prime\prime}\right)\bar{\Phi}_{B^{\prime\prime}}\left(\frac{1}{2}\bm{P}-\bm{k}^{\prime},s^{\prime\prime}\right)V\Phi_{B^{\prime}}\left(\frac{1}{2}\bm{P}+\bm{k},s^{\prime}\right)\Phi_B\left(\frac{1}{2}\bm{P}-\bm{k},s\right),\\
    &&U^{ex}_{B^{\prime\prime\prime}B^{\prime\prime}B^{\prime}B}(\bm{k}^{\prime},\bm{k},s^{\prime\prime\prime},s^{\prime\prime},s^{\prime},s;\bm{P})\nonumber\\
    &\equiv&\bar{\Theta}_{B^{\prime\prime\prime}}\left(\frac{1}{2}\bm{P}+\bm{k}^{\prime},s^{\prime\prime\prime}\right)\bar{\Phi}_{B^{\prime\prime}}\left(\frac{1}{2}\bm{P}-\bm{k}^{\prime},s^{\prime\prime}\right)V\Phi_B\left(\frac{1}{2}\bm{P}-\bm{k},s\right)\Phi_{B^{\prime}}\left(\frac{1}{2}\bm{P}+\bm{k},s^{\prime}\right),\\
    &&W^{dir}_{B^{\prime\prime\prime}B^{\prime\prime}B^{\prime}B}(\bm{k}^{\prime},\bm{k},s^{\prime\prime\prime},s^{\prime\prime},s^{\prime},s;\bm{P})\nonumber\\
    &\equiv&\bar{\Phi}_{B^{\prime\prime\prime}}\left(\frac{1}{2}\bm{P}+\bm{k}^{\prime},s^{\prime\prime\prime}\right)\bar{\Phi}_{B^{\prime\prime}}\left(\frac{1}{2}\bm{P}-\bm{k}^{\prime},s^{\prime\prime}\right)V\Theta_{B^{\prime}}\left(\frac{1}{2}\bm{P}+\bm{k},s^{\prime}\right)\Phi_B\left(\frac{1}{2}\bm{P}-\bm{k},s\right),\\
    &&W^{ex}_{B^{\prime\prime\prime}B^{\prime\prime}B^{\prime}B}(\bm{k}^{\prime},\bm{k},s^{\prime\prime\prime},s^{\prime\prime},s^{\prime},s;\bm{P})\nonumber\\
    &\equiv&\bar{\Phi}_{B^{\prime\prime\prime}}\left(\frac{1}{2}\bm{P}+\bm{k}^{\prime},s^{\prime\prime\prime}\right)\bar{\Phi}_{B^{\prime\prime}}\left(\frac{1}{2}\bm{P}-\bm{k}^{\prime},s^{\prime\prime}\right)V\Phi_B\left(\frac{1}{2}\bm{P}-\bm{k},s\right)\Theta_{B^{\prime}}\left(\frac{1}{2}\bm{P}+\bm{k},s^{\prime}\right),\\
    &&Z^{dir}_{B^{\prime\prime\prime}B^{\prime\prime}B^{\prime}B}(\bm{k}^{\prime},\bm{k},s^{\prime\prime\prime},s^{\prime\prime},s^{\prime},s;\bm{P})\nonumber\\
    &\equiv&\bar{\Theta}_{B^{\prime\prime\prime}}\left(\frac{1}{2}\bm{P}+\bm{k}^{\prime},s^{\prime\prime\prime}\right)\bar{\Phi}_{B^{\prime\prime}}\left(\frac{1}{2}\bm{P}-\bm{k}^{\prime},s^{\prime\prime}\right)V\Theta_{B^{\prime}}\left(\frac{1}{2}\bm{P}+\bm{k},s^{\prime}\right)\Phi_B\left(\frac{1}{2}\bm{P}-\bm{k},s\right),\\
    &&Z^{ex}_{B^{\prime\prime\prime}B^{\prime\prime}B^{\prime}B}(\bm{k}^{\prime},\bm{k},s^{\prime\prime\prime},s^{\prime\prime},s^{\prime},s;\bm{P})\nonumber\\
    &\equiv&\bar{\Theta}_{B^{\prime\prime\prime}}\left(\frac{1}{2}\bm{P}+\bm{k}^{\prime},s^{\prime\prime\prime}\right)\bar{\Phi}_{B^{\prime\prime}}\left(\frac{1}{2}\bm{P}-\bm{k}^{\prime},s^{\prime\prime}\right)V\Phi_B\left(\frac{1}{2}\bm{P}-\bm{k},s\right)\Theta_{B^{\prime}}\left(\frac{1}{2}\bm{P}+\bm{k},s^{\prime}\right),
\end{eqnarray}
where $V$ represents the OBE operator.  In the present calculation, we consider the exchanges of eight mesons ($\sigma,\,\delta,\,\omega,\,\rho,\,\eta,\,\pi,\,K$ and $K^{\ast}$). 

When we solve the ladder-approximated, coupled Bethe-Salpeter equations, Eqs.(\ref{eq_BS_for_T})-(\ref{eq_BS_for_P}), we adopt the standard spinors, Eqs.(\ref{eq_positive_spinor})-(\ref{eq_negative_spinor}), in the nuclear-matter rest frame, and 
do not use the partial-wave decomposition and the helicity spinors (as discussed near Fig.\ref{fig_Schematic_representation_of_momenta}).\footnote{We note that, in Ref.\cite{without_PW}, Sammarruca et al. have also carried out the DBHF calculation without using the partial-wave decomposition.} 
Note that the negative-energy states are included only in the initial and/or final states of the Bethe-Salpeter amplitudes,  namely, in the intermediate states, they do not appear \cite{Poschenrieder,Huber}, because, in the realistic baryon-baryon potentials such as the Bonn potentials \cite{DBHF1}, the negative-energy states are usually excluded. 

To reduce the number of variables and make the present calculation feasible, we here average the azimuthal angle in the spinors, Eqs.(\ref{eq_positive_spinor})-(\ref{eq_negative_spinor}), namely we replace $E^{\ast}_B(1/2\bm{P}\pm\bm{k})$ by the azimuthally averaged effective energy, $\frac{1}{2\pi}\int d\phi E^{\ast}_B(1/2\bm{P}\pm\bm{k})$. 
We have checked that this change does not lead any large numerical error in our final results. 

Given the reaction amplitudes, we can calculate three self-energy components
\begin{eqnarray}
    \Sigma_{\Phi\Phi}^B(k)&\equiv&\bar{\Phi}_B(k,+)\Sigma_B(k)\Phi_B(k,+)\nonumber\\
    &=&\sum_{B^{\prime},s^{\prime}}\int\frac{d^3q}{(2\pi)^3}\frac{\theta(k_{FB^{\prime}}-|\bm{q}|)}{E_{B^{\prime}}^{\ast}(q)}T_{BB^{\prime}BB^{\prime}}\left(\frac{1}{2}(\bm{k}-\bm{q}),\frac{1}{2}(\bm{k}-\bm{q}),+,s^{\prime},+,s^{\prime};\bm{P}\right),\label{eq_SPP}\\
    \Sigma_{\Theta\Phi}^B(k)&\equiv&\bar{\Theta}_B(k,+)\Sigma_B(k)|\Phi_B(k,-)\nonumber\\
    &=&\sum_{B^{\prime},s^{\prime}}\int\frac{d^3q}{(2\pi)^3}\frac{\theta(k_{FB^{\prime}}-|\bm{q}|)}{E_{B^{\prime}}^{\ast}(q)}R_{BB^{\prime}BB^{\prime}}\left(\frac{1}{2}(\bm{k}-\bm{q}),\frac{1}{2}(\bm{k}-\bm{q}),+,s^{\prime},-,s^{\prime};\bm{P}\right),\label{eq_STP}\\
    \Sigma_{\Theta\Theta}^B(k)&\equiv&\bar{\Theta}_B(k,+)\Sigma_B(k)\Theta_B(k,+)\nonumber\\
    &=&\sum_{B^{\prime},s^{\prime}}\int\frac{d^3q}{(2\pi)^3}\frac{\theta(k_{FB^{\prime}}-|\bm{q}|)}{E_{B^{\prime}}^{\ast}(q)}P_{BB^{\prime}BB^{\prime}}\left(\frac{1}{2}(\bm{k}-\bm{q}),\frac{1}{2}(\bm{k}-\bm{q}),+,s^{\prime},+,s^{\prime};\bm{P}\right),\label{eq_STT}
\end{eqnarray}
where $\Sigma_B(k)$ is given by Eq.(\ref{eq_S}), and $k_{FB}$ is the Fermi momentum of baryon $B$.
Substituting Eqs.(\ref{eq_SPP}) - (\ref{eq_STT}) into the following relations 
\begin{eqnarray}
    \Sigma_{\Phi\Phi}^B(k)&=&2M_B^{\ast}(k)\Sigma_B^S(k)-2E^{\ast}_B(k)\Sigma_B^0(k)+2\bm{k}\cdot\bm{k}_B^{\ast}\Sigma_B^V(k), \label{eq_self1} \\
    \Sigma_{\Theta\Phi}^B(k)&=&2|\bm{k}^{\ast}_B|(k)\Sigma_B^0(k)-2|\bm{k}|E_B^{\ast}(k)\Sigma_B^V(k),\\
    \Sigma_{\Theta\Theta}^B(k)&=&-2M_B^{\ast}(k)\Sigma_B^S(k)-2E^{\ast}_B(k)\Sigma_B^0(k)+2\bm{k}\cdot\bm{k}_B^{\ast}\Sigma_B^V(k), \label{eq_self3}
\end{eqnarray}
we can uniquely obtain the self-energies in Eq.(\ref{eq_S}).

The energy density of nuclear matter, ${\cal E}$, is composed of the kinetic and potential energies, which are given in term of $T_{BB^{\prime}BB^{\prime}}$ as
\begin{equation}
    {\cal E}=\sum_B\left(\left<\hat{\cal T}_B\right>+\sum_{B^{\prime}}\left<\hat{\cal V}_{BB^{\prime}}\right>\right),
\end{equation}
with
\begin{eqnarray}
    \left<\hat{\cal T}_B\right>&=&2\int\frac{d^3k}{(2\pi)^3}\left[\frac{M_BM_B^{\ast}(k)+\bm{k}\cdot\bm{k}^{\ast}_B}{E_B^{\ast}(k)}\right],\\
    \left<\hat{\cal V}_{BB^{\prime}}\right>&=&\int\frac{d^3p}{(2\pi)^3}\int\frac{d^3P}{(2\pi)^3}\frac{\theta(k_{F_B}-|\frac{1}{2}\bm{P}+\bm{p}|)}{2E_B^{\ast}(\frac{1}{2}\bm{P}+\bm{p})}\frac{\theta(k_{F_{B^{\prime}}}-|\frac{1}{2}\bm{P}-\bm{p}|)}{E_{B^{\prime}}^{\ast}(\frac{1}{2}\bm{P}-\bm{p})}\nonumber\\
    &&\times\sum_{s,s^{\prime}}T_{BB^{\prime}BB^{\prime}}\left(\bm{p},\bm{p},s,s^{\prime},s,s^{\prime};\bm{P}\right),
\end{eqnarray}
where we omit the retardation effect by neglecting terms being proportional to the energy transfer between two baryons. 
We can calculate pressure of nuclear matter in the similar manner \cite{DBHF3}. 

\section{Results \label{sec:results}}
We solve the Bethe-Salpeter equations, Eqs.(\ref{eq_BS_for_T})-(\ref{eq_BS_for_P}), in the nuclear-matter rest frame by using the usual spinors, Eqs.(\ref{eq_positive_spinor})-(\ref{eq_negative_spinor}). Then, we carry out the baryon self-energies, the energy density and pressure of nuclear matter or neutron-star matter.  In the present calculation, we use the Bonn potentials \cite{DBHF1}, and try to consider not only nucleon but also the degrees of freedom of hyperons.  We first study the properties of symmetric nuclear matter, and then consider neutron-star matter, where the possible, maximum mass of neutron stars is calculated. 

\subsection{Symmetric nuclear matter \label{sec:symmetric_nuclear_matter}}
In this section, we study symmetric nuclear matter, where matter consists of only nucleons interacting through the exchanges of $\sigma,\,\delta,\,\omega,\,\rho,\,\eta$ and $\pi$ mesons. 

\begin{figure}[htbp]
 \begin{center}
    \includegraphics[width=240pt,keepaspectratio,clip,angle=270]{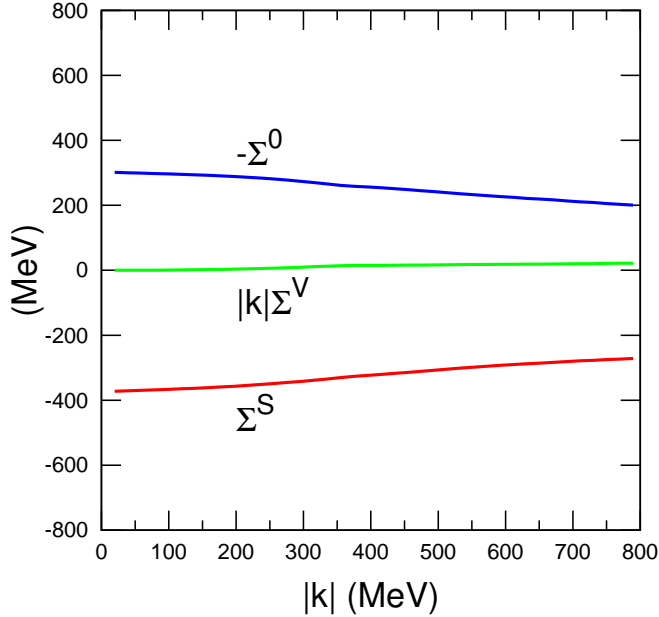}
 \end{center}
 \caption{(Color online) Momentum dependence of the nucleon self-energies in symmetric nuclear matter at $k_{F}^0=1.34\,\mathrm{fm}^{-1}$.  The result is for the Bonn A potential.}
 \label{fig_momentum_dependence_of_self_energy}
\end{figure}

We first examine how the three self-energies depend on the nucleon momentum in matter.  
Supposing that the momentum at the saturation point (or the normal nuclear density), $n_B^0$, is given by the empirical value, $k_{F}^0=1.34\,\mathrm{fm}^{-1}$ \cite{Fuchs2}, the self-energies can be calculated through Eqs.(\ref{eq_self1})-(\ref{eq_self3}). 
In this calculation, they satisfy Eqs.(\ref{eq_MsB})-(\ref{eq_EsB}) at each $|\bm{k}|$, namely the momentum dependence of each self-energy is fully taken into account \cite{Poschenrieder,Huber}.  We show the momentum dependence of each self-energy in Fig.\ref{fig_momentum_dependence_of_self_energy}.  
As seen in the figure, the momentum dependence is not strong.  In particular, 
the space component of the self-energy, $\Sigma^V$, is almost constant, and the absolute value is substantially smaller than those of other components, $\Sigma^S$ and $\Sigma^0$.

This fact implies that the assumption often used in the conventional DBHF approach, namely $\Sigma^V$ is assumed to be negligible (as discussed in section \ref{sec:intro}), is reasonable around the saturation density. 
We also comment that the present self-energies are very similar to those given in Fig.8 of Ref.\cite{Fuchs2}, in which the calculation has been performed in the subtraction scheme with the pseudoscalar (ps) representation and only the positive-energy states of nucleon are taken into account \cite{Fuchs2}. 
This may also infer that the effect of negative-energy states is limited around $n_B^0$ (see also the discussion in section \ref{sec:formalism}), as long as one chooses the subtraction scheme with the ps representation.\footnote{In Fig.\ref{fig_momentum_dependence_of_self_energy}, we can find that the present self-energies are slightly different from those in Fig.8 of Ref.\cite{Fuchs2}. This small difference may be caused by the effect of negative-energy states.}

Consequently, up to around $n_B^0$, it is certainly expected that the space part of the self-energy and the effect of negative-energy states can be ignored (although, in such case, the relationship between the in-medium reaction matrices and the self-energies is not clear).  

\begin{figure}[htbp]
 \begin{center}
    \includegraphics[width=250pt,keepaspectratio,clip,angle=270]{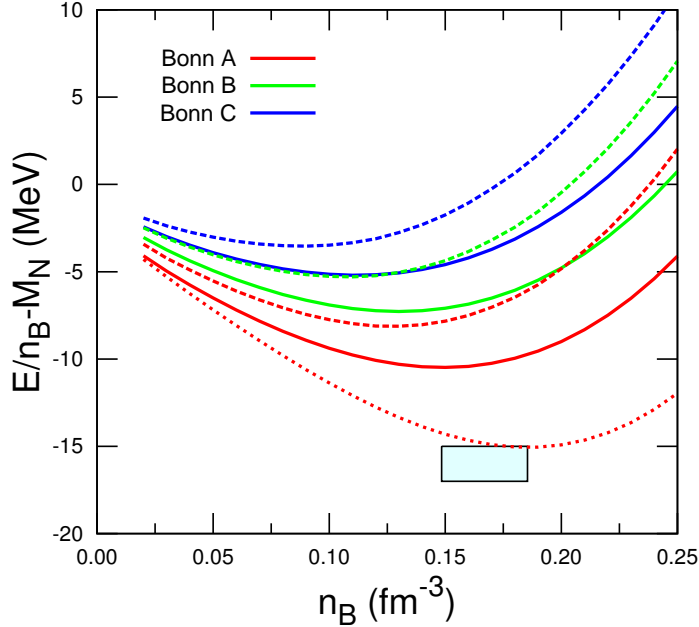}
 \end{center}
 \caption{(Color online) Binding energy per particle in symmetric nuclear matter as a function of $n_B$.
The calculations have been performed using the Bonn potentials A, B and C. 
The dashed (solid) lines represent the saturation curves in the DBHF$^{(1)}$ (DBHF$^{(2)}$) calculation.  The (red) dotted line is for the result with the rescaled coupling, $g_{NN\sigma}^{\ast}$ (see Eq.(\ref{eq:gBBs})), in the DBHF$^{(2^{\ast})}$ approach. The empirical values for the saturation density and binding energy lie within the shaded rectangular region. For details, see the text.}
 \label{fig_saturation}
\end{figure}

In Fig.\ref{fig_saturation}, we present the binding energy per particle in symmetric nuclear matter as a function of the total baryon density, $n_B$. Table \ref{tab:matter_property} also shows several properties of symmetric nuclear matter at $n_B^0$.
\begin{table}
\caption{\label{tab:matter_property}
Calculated properties of symmetric nuclear matter at the saturation point, $n_B^0$.  We use the Bonn potentials A, B and C. In the first column, for example, the result of Bonn A is labeled as DBHF$^{(1)}$(A), etc. 
The values of the binding energy per particle, ${\cal E}/n_B^0-M_N$, the incompressibility, $K$, the symmetry energy, $S$,  and the slope parameter, $L$, are in MeV, and $n_B^0$, is in fm$^{-3}$. For details, see the text.}
\begin{ruledtabular}
\begin{tabular}{ccccccc}
Case&$n_B^0$&${\cal E}/n_B^0-M_N$&$K$&$S$&$L$\\
\hline
DBHF$^{(1)}$(A)&0.124&-8.1&147&24.4&60.7\\
DBHF$^{(1)}$(B)&0.109&-5.3&103&19.4&47.2\\
DBHF$^{(1)}$(C)&0.087&-3.5&53&14.6&31.6\\
\hline
DBHF$^{(2)}$(A)&0.149&-10.5&204&28.8&78.6\\
DBHF$^{(2)}$(B)&0.130&-7.3&133&22.7&58.2\\
DBHF$^{(2)}$(C)&0.112&-5.2&87&18.0&42.2\\
\hline
DBHF$^{(2^{\ast})}$(A)&0.186&-15.1&402&36.8&117.0\\
\end{tabular}
\end{ruledtabular}
\end{table}
In the figure and table, DBHF$^{(1)}$ represents the result of the case where the momentum dependence of each self-energy is fully taken into account (as in Fig.\ref{fig_momentum_dependence_of_self_energy}) \cite{Poschenrieder,Huber}.  
In contrast, DBHF$^{(2)}$ denotes the result in the conventional ``reference spectrum'' approximation \cite{DBHF1,DBHF2,DBHF3,Fuchs1,Fuchs2,Fuchs3,Dalen,without_PW,Jong}, where the values of self-energies are frozen at some reference momentum.\footnote{The ``reference spectrum'' approximation may conventionally imply the neglect of the space part of self-energy as well. However, in the present paper, we do not use this terminology in that sense, because we explicitly include $\Sigma^V$ in the calculation.}
In this calculation, the reference momentum is chosen to be the Fermi momentum, $k_F$, at each $n_B$. 

In our previous calculation \cite{DBHF3}, where we have assumed the subtraction scheme with the ps representation, the results for the Bonn A and B potentials are in the rectangular region of the empirical saturation condition. 
However, in the present DBHF approach, the saturation point in each case follows a new ``Coester lines'', which is slightly  shifted upwards, and thus the binding per particle becomes shallower than the previous ones.  We here comment on this fact: in relativistic dynamics, the binding energy ($\sim 15.5$ MeV empirically) generally results from a sensitive cancellation between the two self-energies, $\Sigma^S$ and $\Sigma^0$, where each is of the order of several hundred MeV. Thus, the binding is very sensitive to the change of the self-energies.  As seen in Fig.\ref{fig_momentum_dependence_of_self_energy}, although the present self-energies are very close to the calculated ones previously in Ref.\cite{DBHF3}, the small difference between them has affected the value of binding at the saturation point. 

Now we try to adjust the binding in the present method by assuming that only the nucleon-$\sigma$ coupling constant varies as
\begin{equation}
    g_{NN\sigma}^{\ast} = g_{NN\sigma}\left(1-\beta \frac{\Sigma^S_N}{M_N} \right) , \label{eq:gBBs}
\end{equation}
where $\beta$ is a constant parameter.  Here, the coupling constant in free space, $g_{NN\sigma}$, is artificially rescaled by a factor determined by the scalar self-energies of nucleon ($N =$ proton or neutron).  Note that, for $\beta=0$ or $n_B=0$, the vacuum value, $g_{NN\sigma}$, is recovered.  

Using this modified coupling constant, we calculate the properties of symmetric nuclear matter within DBHF$^{(2)}$(A) with $\beta=0.06$, which is labeled as DBHF$^{(2^{\ast})}$, and show the result in Fig.\ref{fig_saturation} and Table \ref{tab:matter_property}.  In this calculation, although $g_{NN\sigma}$ is actually enhanced by only about 2\% at $n_B^0$, the saturation point considerably moves toward the empirical region.  However, the incompressibility and the slope parameter seem to be overestimated in this simple modification. By varying not only the coupling between nucleon and $\sigma$ but also other couplings and cutoff parameters in the form factors \cite{Jong}, it may be possible to find a better result in the present method.\footnote{Because our aim is to study the properties of dense neutron-star matter including  hyperons, we do not perform such fine-tuned calculations in this paper.} 

\begin{figure}[htbp]
 \begin{center}
    \includegraphics[width=230pt,keepaspectratio,clip,angle=270]{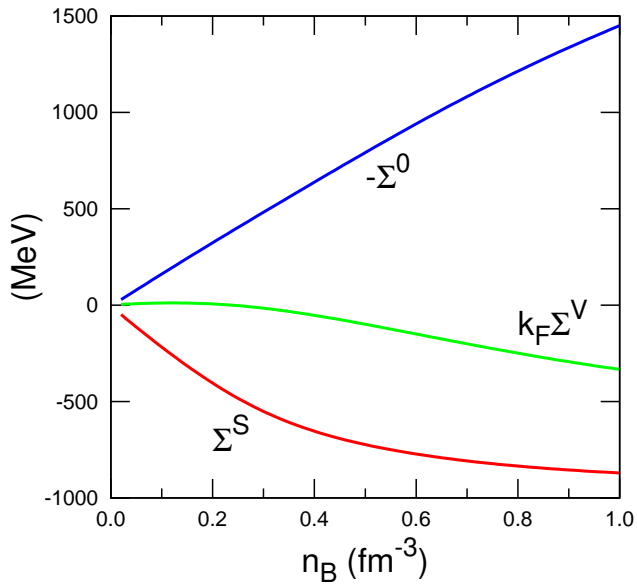}
 \end{center}
 \caption{(Color online) Self-energies in symmetric nuclear mater. The calculation is performed within DBHF$^{(2)}$(A), where the reference momentum is taken to be $k_F$ at each $n_B$.}
 \label{fig_self_energy_without_momentum_dependence}
\end{figure}

As mentioned in the discussion of Fig.\ref{fig_momentum_dependence_of_self_energy}, we have confirmed that the space part of the self-energy, $\Sigma^V$, 
can be neglected around $n_B^0$.  However, it is necessary to examine if such assumption could work well at higher densities.  In Fig.\ref{fig_self_energy_without_momentum_dependence}, we present the density dependence of the self-energies up to $n_B = 1.0\,\mathrm{fm}^{-3}$ in symmetric nuclear matter. 
Because the computation of the momentum-dependent self-energies up to high density is a hard task, instead of the full calculation, we here calculate them within DBHF$^{(2)}$(A).
Indeed, the absolute value of $\Sigma^V$ is very small up to $2n_B^0$, but, with increasing $n_B$, 
it grows rapidly and reaches about $0.7$ at $n_B = 1.0\,\mathrm{fm}^{-3}$.  Furthermore, we note that the space part of the neutron self-energy becomes more enhanced in 
neutron-rich matter than in symmetric matter. 
Therefore, it is unavoidable to consider the space part of the self-energy explicitly in extremely dense, isospin-asymmetric matter.

\subsection{Neutron stars}
\subsubsection{Neutron stars without hyperons\label{protoneutron_stars}}
\begin{figure}[htbp]
 \begin{center}
    \includegraphics[width=250pt,keepaspectratio,clip,angle=270]{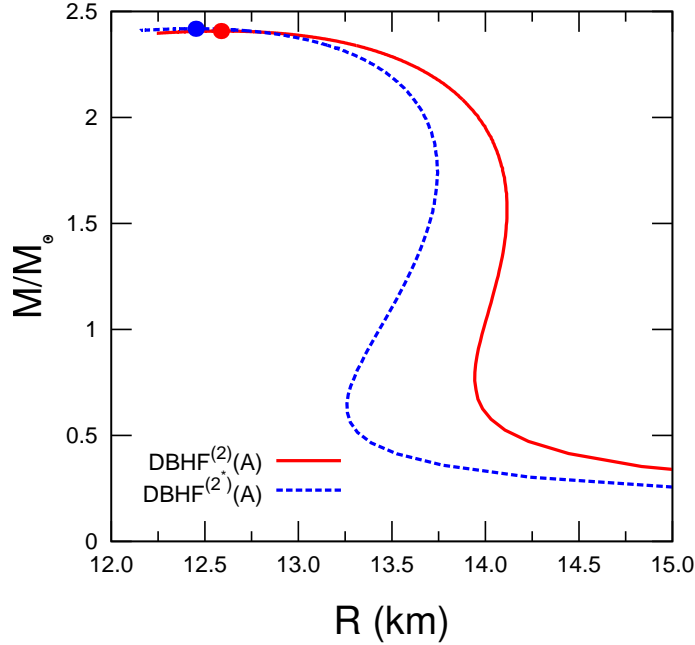}
 \end{center}
 \caption{(Color online) Neutron-star mass versus radius.
The solid (dashed) line is for the result with DBHF$^{(2)}$(A) (DBHF$^{(2^{\ast})}$(A)).
The dot on the line represents the maximum mass (see also Table \ref{tab:NS_Nl_property}).}
 \label{fig_NS_Nl}
\end{figure}

As we have discussed in the previous section, the two approaches, DBHF$^{(2)}$(A) and DBHF$^{(2^{\ast})}$(A), have relatively well explained the properties of symmetric nuclear matter around $n_B^0$, although they have produced the different binding energies and incompressibilities.  
Thus, using the Tolman-Oppenheimer-Volkoff (TOV) equation \cite{TOV} with the BPS model \cite{BPS} for the EoS in the crust region of neutron stars, we here calculate the neutron-star mass as a function of the radius within the degrees of freedom of nucleon, electron and muon, and see how the maximum mass of neutron stars is affected by the difference between the two approaches.  The calculation is performed under the conditions of charge neutrality and $\beta$-equilibrium in weak interaction. The result is shown in Fig.\ref{fig_NS_Nl} and in Table \ref{tab:NS_Nl_property}.\footnote{ To reduce the calculation time, we have used the angle-averaged self-energy. For details, see Ref.\cite{DBHF3}.} 
\begin{table}[htbp]
\caption{\label{tab:NS_Nl_property}
Neutron-star radius, $R_{max}$ (in km), the central density, $n_c$ (in fm$^{-3}$), and the ratio of the maximum neutron-star mass to the solar mass, $M_{max}/M_{\odot}$.}
\begin{ruledtabular}
\begin{tabular}{cccc}
Case&$R_{max}$&$n_c$&$M_{max}/M_{\odot}$\\
\hline
DBHF$^{(2)}$(A)&12.6&0.79&2.41\\
DBHF$^{(2^{\ast})}$(A)&12.5&0.80&2.42\\
\end{tabular}
\end{ruledtabular}
\end{table}

We can see that the two approaches produce the similar results around the maximum mass, whereas, under the maximum mass, the mass-radius relations in the two approaches are somewhat different from each other. 
The curve of DBHF$^{(2^{\ast})}$(A), where the attractive force due to the $\sigma$ exchange is enhanced in matter (see Eq.(\ref{eq:gBBs})), suggests the smaller radius than in that of DBHF$^{(2)}$(A).  We, however, note that, in this region, the EoS for the crust part \cite{BPS} considerably acts upon the mass-radius relation as well. 
Therefore, as far as we primarily concern the maximum mass of neutron stars, it may not matter much whether the saturation properties of nuclear matter at $n_B^0$ are completely reproduced or not.

\subsubsection{Inclusion of hyperons \label{including_hyperons}}
In the following calculations, we choose the method of DBHF$^{(2)}$(A).

\begin{figure}[htbp]
 \begin{center}
    \includegraphics[width=380pt,keepaspectratio,clip]{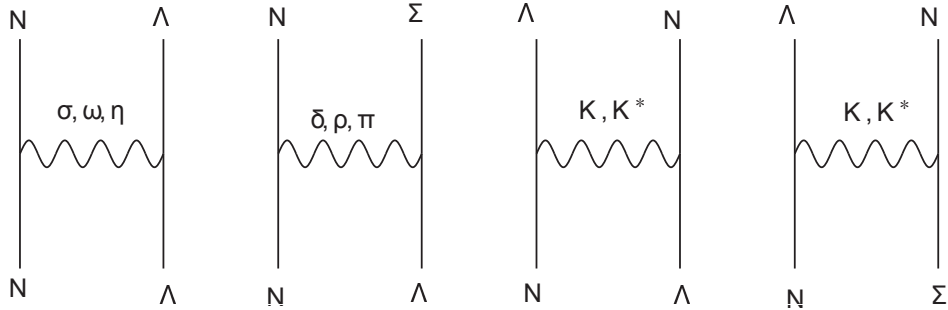}
 \end{center}
 \caption{One-boson-exchange processes for the $N\Lambda$ system, in which the scalar ($\sigma,\,\delta$), vector ($\omega,\,\rho,\,K^{\ast}$) and pseudoscalar ($\eta,\,\pi,\,K$) mesons take part.  The $\Sigma$ hyperon partly contributes to the system as well. 
}
 \label{fig_Nucleon_Lambda_interactions}
\end{figure}

We now consider the degrees of freedom of hyperons as well as nucleon.  In this case, the coupled Bethe-Salpeter equations are very complex.
For example, the $N$-$\Lambda$ system includes four OBE processes shown in 
Fig.\ref{fig_Nucleon_Lambda_interactions}.  
In RMFT, only the left diagram is considered, while, in the present DBHF calculation, we have to take into account the other three diagrams as well, where the baryon-exchange and baryon-transition processes are included. 

In this calculation, the force between two baryons is generated by the exchanges of eight mesons ($\sigma,\,\delta,\,\omega,\,\rho,\,\eta,\,\pi,\,K$ and $K^{\ast}$). 
It is then well known that the tensor parts of the $\pi$- and $\rho$-exchange potentials are mostly canceled each other, and that 
it is true even in the $K$ and $K^{\ast}$-exchange potentials \cite{Holzenkamp}. It is thus important to treat them as a pairwise.  Using the interactions due to those meson exchanges, we self-consistently calculate the EoS including hyperons. 

Because the present calculation is very heavy and demands an enormous calculation time, 
we first study neutron stars with hyperons using the interactions which do not induce the baryon-exchange and baryon-transition processes, namely only the diagrams like the left one in Fig.\ref{fig_Nucleon_Lambda_interactions} are considered.\footnote{Note that, in this case, the $K$ and $K^\ast$ mesons are excluded because they always induce the baryon-exchange and baryon-transition processes.} 
Then, we perform the calculations for two cases: one includes $e^-$, $\mu^-$ and five baryons (neutrons, protons, $\Lambda$, $\Sigma^-$, $\Xi^-$),\footnote{
From among the members of the $\Sigma$ and $\Xi$ hyperons in SU(3) flavor symmetry, we select only the $\Sigma^-$ and $\Xi^-$.  The reason is because, from the viewpoint of electric charge, it is expected that they can appear easier in matter rather than the other members ($\Sigma^+$, $\Sigma^0$ and $\Xi^0$).} which we call NlY5, and the other the leptons and eight baryons (neutrons, protons, $\Lambda$, $\Sigma^-$, $\Sigma^0$, $\Sigma^+$, $\Xi^-$, $\Xi^0$), which we call NlY8. 
After that, we challenge the calculation including the baryon-exchange and baryon-transition processes.  

Using the experimental data of nucleon-hyperon scattering, the hyperon-meson coupling constants have been studied by several groups \cite{Holzenkamp,Haidenbauer,Rijken}.
However, due to poor experimental accuracy, the coupling constants cannot be determined without large ambiguities. 
Thus, in the present calculation, we determine them with SU(6) symmetry \cite{SU6}: 
\begin{eqnarray}
    g_{NN\omega}=\frac{3}{2}g_{\Lambda\Lambda\omega}=\frac{3}{2}g_{\Sigma\Sigma\omega}=3g_{\Xi\Xi\omega},\\
    g_{NN\rho}=\frac{1}{2}g_{\Sigma\Sigma\rho}=g_{\Xi\Xi\rho},\,g_{\Lambda\Lambda\rho}=0,\\
    g_{NN\delta}=\frac{1}{2}g_{\Sigma\Sigma\delta}=g_{\Xi\Xi\delta},\,g_{\Lambda\Lambda\delta}=0,\\
    g_{NN\pi}=-\frac{5}{2\sqrt{3}}g_{\Lambda\Lambda\eta}=\frac{5}{2\sqrt{3}}g_{\Sigma\Sigma\eta}=-\frac{5}{3\sqrt{3}}g_{\Xi\Xi\eta},\\
    g_{NN\pi}=\frac{5}{4}g_{\Sigma\Sigma\pi}=-5g_{\Xi\Xi\pi},\,g_{\Lambda\Lambda\pi}=0,\\
    g_{\Lambda\Sigma\rho}=g_{\Lambda\Sigma\delta}=0,\,g_{\Lambda\Sigma\pi}=\frac{2\sqrt{3}}{5}g_{NN\pi},\\
    g_{NN\pi}=-\frac{5}{3\sqrt{3}}g_{N\Lambda K}=5g_{N\Sigma K}=\frac{5}{\sqrt{3}}g_{\Lambda\Xi K}=-g_{\Sigma\Xi K},\\
    g_{NN\rho}=-\frac{1}{\sqrt{3}}g_{N\Lambda K^{\ast}}=-g_{N\Sigma K^{\ast}}=\frac{1}{\sqrt{3}}g_{\Lambda\Xi K^{\ast}}=-g_{\Sigma\Xi K^{\ast}},
\end{eqnarray}
and
\begin{eqnarray}
    \kappa_{\Lambda\Lambda\omega}\equiv\frac{f_{\Lambda\Lambda\omega}}{g_{\Lambda\Lambda\omega}}=-1,\,\kappa_{\Sigma\Sigma\omega}\equiv\frac{f_{\Sigma\Sigma\omega}}{g_{\Sigma\Sigma\omega}}=-\frac{3}{5}+\frac{2}{5}\kappa_{NN\rho},\\
    \kappa_{\Xi\Xi\omega}\equiv\frac{f_{\Xi\Xi\omega}}{g_{\Xi\Xi\omega}}=-\frac{6}{5}-\frac{1}{5}\kappa_{NN\rho},\,\kappa_{\Lambda\Lambda\rho}\equiv\frac{f_{\Lambda\Lambda\rho}}{g_{\Lambda\Lambda\rho}}=0,\\
    \kappa_{\Sigma\Sigma\rho}\equiv\frac{f_{\Sigma\Sigma\rho}}{g_{\Sigma\Sigma\rho}}=-\frac{3}{5}+\frac{2}{5}\kappa_{NN\rho},\,\kappa_{\Xi\Xi\rho}\equiv\frac{f_{\Xi\Xi\rho}}{g_{\Xi\Xi\rho}}=-\frac{6}{5}-\frac{1}{5}\kappa_{NN\rho},\\
    \kappa_{N\Lambda K^{\ast}}\equiv\frac{f_{N\Lambda K^{\ast}}}{g_{N\Lambda K^{\ast}}}=-\frac{2}{5}+\frac{3}{5}\kappa_{NN\rho},\,\kappa_{N\Sigma K^{\ast}}\equiv\frac{f_{N\Sigma K^{\ast}}}{g_{N\Sigma K^{\ast}}}=-\frac{6}{5}-\frac{1}{5}\kappa_{NN\rho},\\
    \kappa_{\Lambda\Xi K^{\ast}}\equiv\frac{f_{\Lambda\Xi K^{\ast}}}{g_{\Lambda\Xi K^{\ast}}}=-\frac{4}{5}+\frac{1}{5}\kappa_{NN\rho},\,\kappa_{\Sigma\Xi K^{\ast}}\equiv\frac{f_{\Sigma\Xi K^{\ast}}}{g_{\Sigma\Xi K^{\ast}}}=\kappa_{NN\rho},\\
    f_{\Lambda\Sigma\rho}=\frac{2\sqrt{3}}{5}(1+\kappa_{NN\rho})g_{NN\rho}, 
\end{eqnarray}
where $\kappa$ is defined by the ratio of the tensor to vector coupling constants of a vector meson.
The coupling constant of a pseudoscalar meson with mass, $m_{ps}$, in the pseudoscalar coupling, $g_{BB^{\prime}ps}$, is related to that in the pseudovector coupling, $f_{BB^{\prime}ps}$, as 
\begin{equation}
    f_{BB^{\prime}ps}=\frac{m_{ps}}{M_B+M_{B^{\prime}}}g_{BB^{\prime}ps}.
\end{equation}

In addition, it is necessary to determine the coupling constant for the hyperon-$\sigma$ interaction.  The recent analyses of hypernuclei and hyperon production reactions suggest that the $\Lambda,\,\Sigma^{-}$ and $\Xi^{-}$ respectively feel the optical potential, $U_{\Lambda^{-},\Sigma^{-},\Xi^{-}}\simeq-27,\,+30,\,-15$ MeV, in a nuclear medium \cite{hyppot}.
We thus fix the hyperon-$\sigma$ coupling constants, $g_{BB\sigma}$, so as to reproduce these potential depths at $n_B^0$, using the Schr\"{o}edinger-equivalent, baryon optical potential given by
\begin{equation}
    U_B(k) = \Sigma^S_B(k)-\frac{\Sigma^0_B(k)}{M_B}(E^{\ast}_B(k)-\Sigma^0_B(k))+\frac{1}{2M_B}(\Sigma_B^{S\,2}(k)-\Sigma^{0\,2}_B(k)).
\end{equation}

Furthermore, for nonstrange mesons, a cutoff parameter in the form factor at the hyperon-meson vertex, $\Lambda_{BB^{\prime}M}$, is assumed to be the same value as in the nucleon-meson form factor, while, for strange mesons, a cutoff parameter, $\Lambda_{BB^{\prime}K}$ ($\Lambda_{BB^{\prime}K^{\ast}}$), is taken to be  the average value of $\Lambda_{BB^{\prime}\eta}$ and $\Lambda_{BB^{\prime}\pi}$ ($\Lambda_{BB^{\prime}\omega}$ and $\Lambda_{BB^{\prime}\rho}$).  

\begin{figure}[htbp]
 \begin{minipage}{.48\textwidth}
    \includegraphics[width=165pt,keepaspectratio,clip,angle=270]{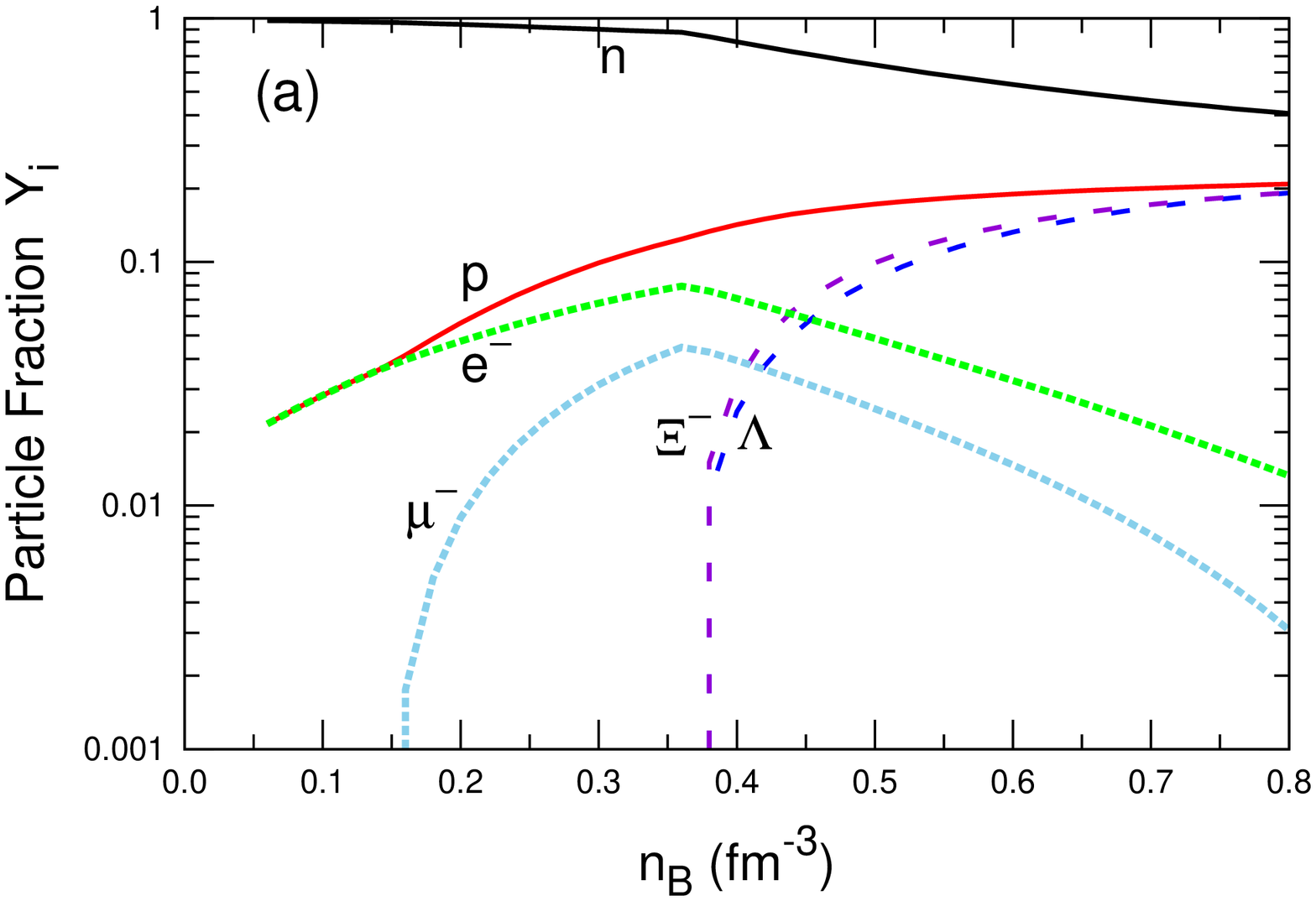}
 \end{minipage}
 \begin{minipage}{.48\textwidth}
    \includegraphics[width=165pt,keepaspectratio,clip,angle=270]{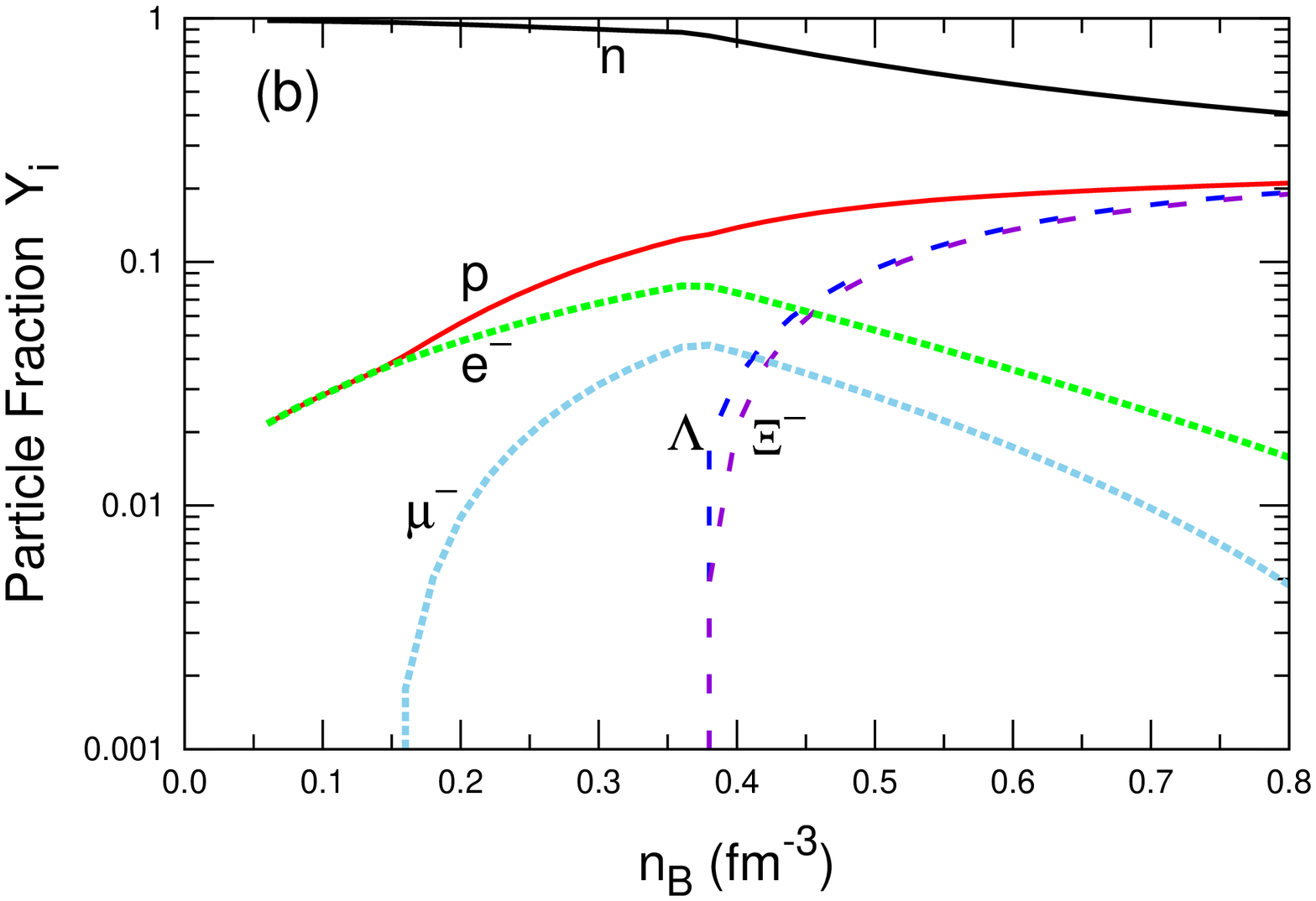}
 \end{minipage}
 \caption{(Color online) Particle fractions, $Y_i$, for (a) NlY5 and (b) NlY8.}
 \label{fig_NS_NlY5vs8}
\end{figure}

Now we are in a position to show our results. 
In Fig.\ref{fig_NS_NlY5vs8}, we present the particle fractions, $Y_i\,(i=n,p,\cdots)$, in neutron-star matter. 
As seen in the figure, both the results are very similar to each other, and show that the $\Lambda$ and $\Xi^-$ first appear around $n_B\simeq0.38\,\mathrm{fm}^{-3}$.
This tendency of the particle fractions is also close to that calculated in RH or relativistic Hartree-Fock (RHF) approximation \cite{ApJ,miyatsu}, although the present mass-radius relation is quite different from that in the RH or RHF calculation (see below for further details). 

Because the difference between the results in NlY5 and NlY8 is expected to be very small, we proceed to the next  calculation, where the $K$ and $K^{\ast}$ mesons are considered and thus the baryon-exchange and baryon-transition processes take place, within the NlY5 scheme.  We call this NlY5KK$^{\ast}$.  

\begin{figure}[htbp]
 \begin{center}
    \includegraphics[width=350pt,keepaspectratio,clip,angle=270]{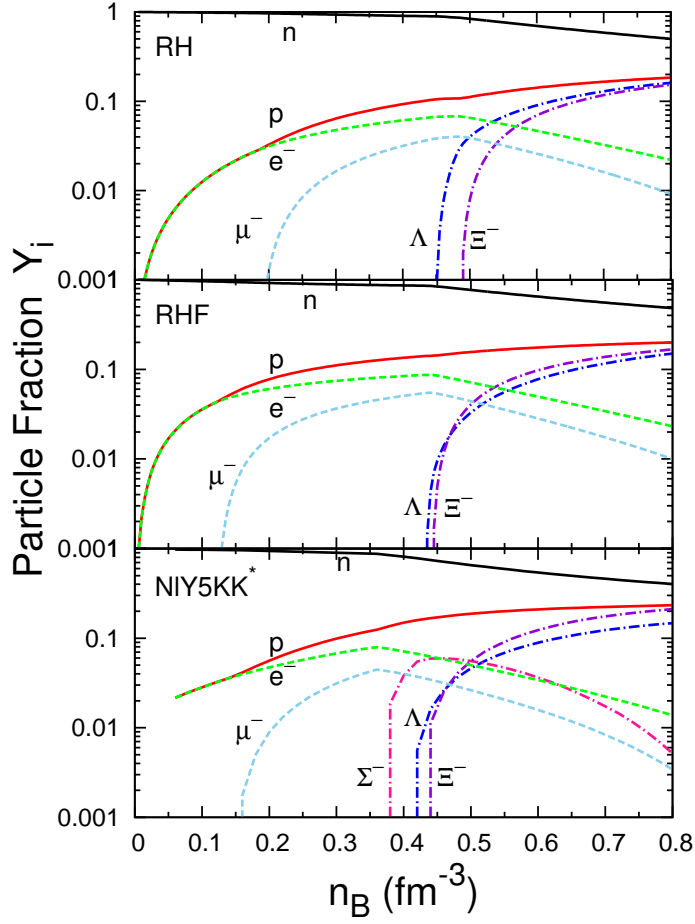}
 \end{center}
 \caption{(Color online) Particle fractions, $Y_i$, in cases of RH, RHF and NlY5KK$^{\ast}$. The calculations in RH and RHF approximations are explained in Ref.\cite{ApJ}.}
 \label{fig_NS_NlY5KK}
\end{figure}

In Fig.\ref{fig_NS_NlY5KK}, the particle fractions in the calculation of NlY5KK$^{\ast}$, together with the results in RH and RHF approximations with SU(6) symmetry, are displayed.  
Including the baryon-exchange and baryon-transition processes, the $\Sigma^-$ first appears at $n_B\simeq0.38\,\mathrm{fm}^{-3}$, but its amount is rapidly reduced with increasing $n_B$.  Furthermore, comparing with the result in NlY5, the threshold densities for the $\Lambda$ and $\Xi^{-}$ in NlY5KK$^{\ast}$ are pushed toward higher densities;  $n_B\simeq0.42\,\mathrm{fm}^{-3}$ for the $\Lambda$ and $n_B\simeq0.44\,\mathrm{fm}^{-3}$ for the $\Xi^{-}$.  In the three cases, the results are similar to one another, except for the appearance of $\Sigma^-$ in NlY5KK$^{\ast}$.  
\begin{figure}[htbp]
 \begin{center}
    \includegraphics[width=250pt,keepaspectratio,clip,angle=270]{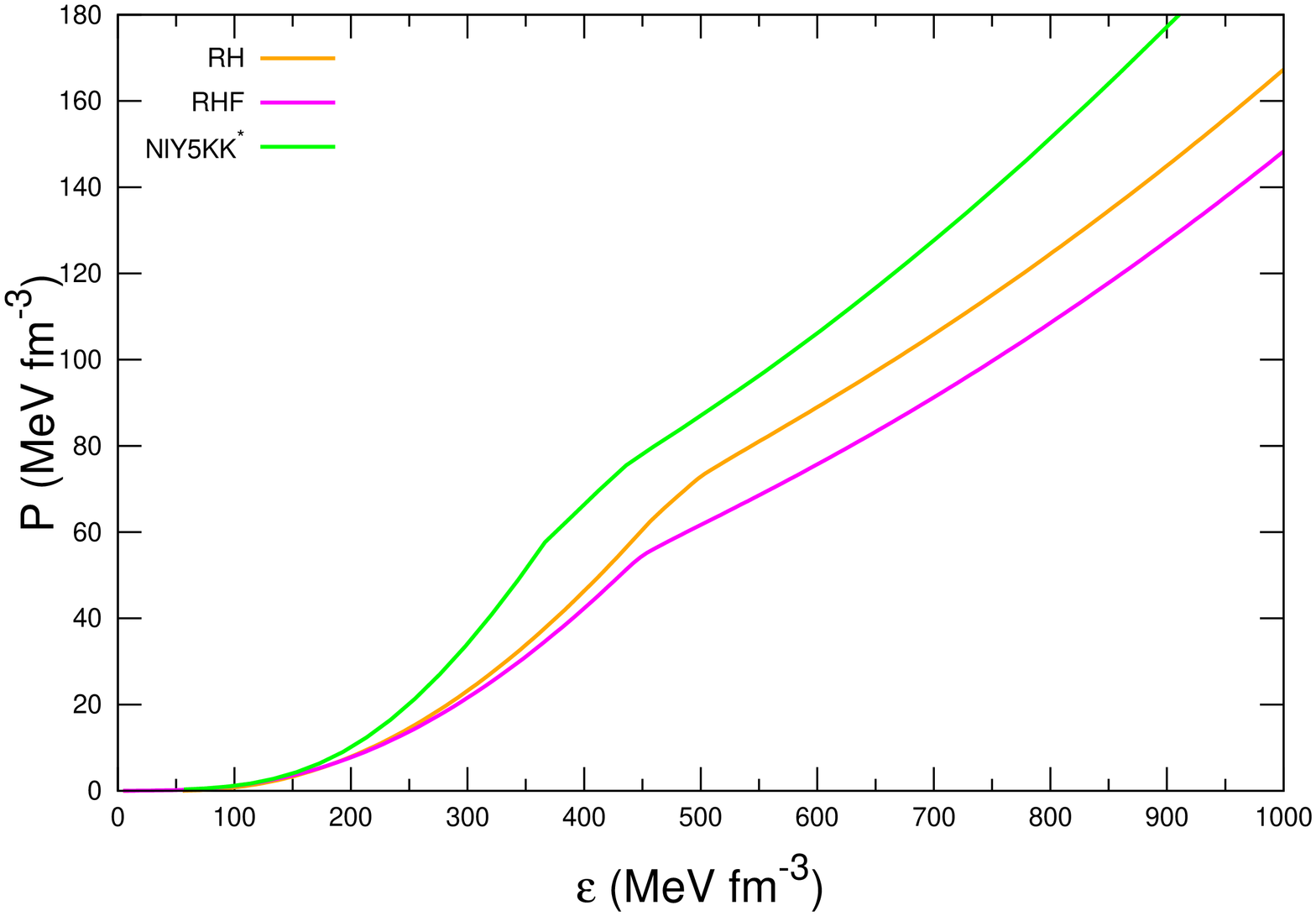}
 \end{center}
 \caption{(Color online) Equations of state in cases of RH, RHF and NlY5KK$^{\ast}$. }
 \label{fig_NS_EoS}
\end{figure}
\begin{figure}[htbp]
 \begin{center}
    \includegraphics[width=310pt,keepaspectratio,clip,angle=270]{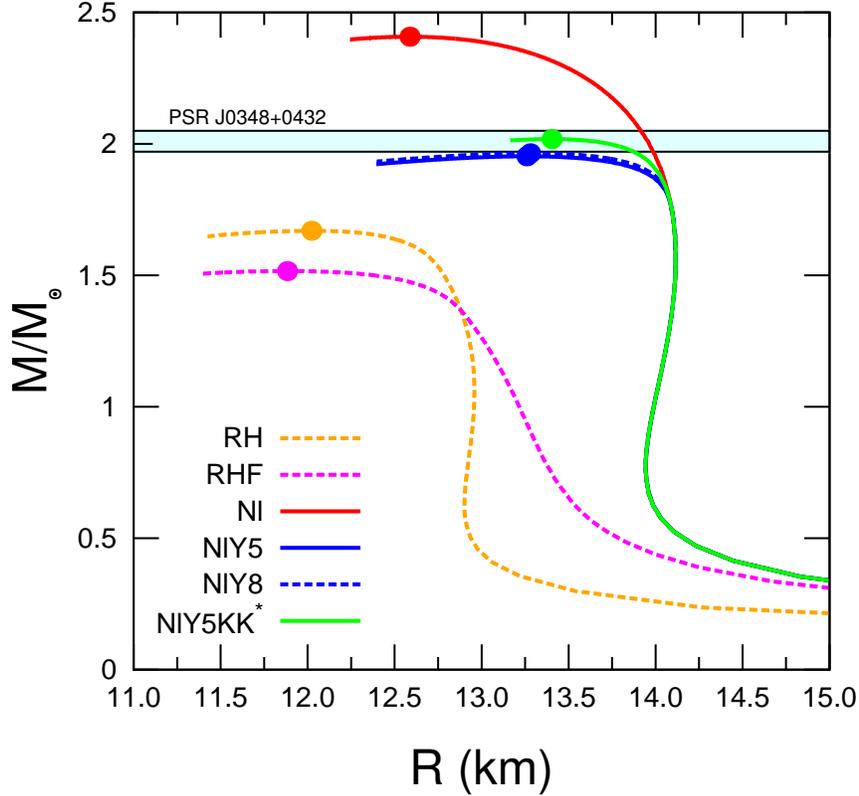}
 \end{center}
 \caption{(Color online) Mass-radius relations for neutron stars in various cases.
The dot on each line represents the maximum mass (see also Table \ref{tab:NS_NlY_property}). The shaded area represents the observed constraint by the pulsar J0348+0432.}
 \label{fig_NS_NlY}
\end{figure}
\begin{table}
\caption{\label{tab:NS_NlY_property}
Neutron-star radius, $R_{max}$ (in km), the central density, $n_c$ (in fm$^{-3}$), and the ratio of the maximum neutron-star mass to the solar mass, $M_{max}/M_{\odot}$. The case of Nl corresponds to the result of DBHF$^{(2)}$(A) in Table \ref{tab:NS_Nl_property}.}
\begin{ruledtabular}
\begin{tabular}{cccc}
Case&$R_{max}$&$n_c$&$M_{max}/M_{\odot}$\\
\hline
Nl&12.6&0.79&2.41\\
NlY5&13.3&0.74&1.95\\
NlY8&13.3&0.73&1.96\\
NlY5KK$^{\ast}$&13.4&0.71&2.02\\\hline
RH&12.0&0.89&1.67\\
RHF&11.9&0.94&1.52\\
\end{tabular}
\end{ruledtabular}
\end{table}
However, we see the large difference among the equations of sate in three cases, which is shown in Fig.\ref{fig_NS_EoS}. 
At high densities, the EoS in the DBHF calculation is much harder than in the RH and RHF calculations.\footnote{
In Refs.\cite{ApJ,miyatsu}, we have reported that, in the case where the coupling constants are determined in SU(3) flavor symmetry and the quark degrees of freedom inside baryons are considered \cite{saito}, the RHF calculation can also produce a relatively hard EoS.}

In Fig.\ref{fig_NS_NlY} and Table \ref{tab:NS_NlY_property}, our results are summarized, and they show the mass-radius relations and the properties of neutron stars at the maximum mass.  
Corresponding to the fact that the hard EoS is obtained in the DBHF case, 
we can find that the predicted maximum masses in the DBHF calculations are very close to the observed one, $1.97\pm0.04M_{\odot}$ (J1614-2230).  In particular, the mass in the case of NlY5KK$^{\ast}$ reaches $2.02\,M_{\odot}$, which is also consistent with the constraint, $2.01\pm0.04M_{\odot}$, provided by the pulsar (J0348+0432).

In RH or RHF approximation with SU(6) symmetry, because the appearance of hyperons in matter inevitably softens the EoS, the maximum neutron-star mass becomes seriously diminished \cite{ApJ}.  However, in the DBHF calculation, although hyperons appear as in cases of RH and RHF, the maximum mass can be much heavier than in the mean-field case. 
The reason for the difference is that the DBHF calculation involves the strong density dependence of the in-medium baryon-baryon scattering amplitude, which is mainly caused by the Pauli exclusion principle, the short-range baryon-baryon correlations and many-body forces in matter. 
This density dependence is thus significant in extremely dense matter like neutron stars.

\section{Summary and conclusions\label{sec:summary_and_conclusion}}

Our aim in the present paper is to apply the Dirac-Brueckner-Hartree-Fock approach to the equation of state for extremely dense, nuclear matter, in which the degrees of freedom of hyperons as well as nucleon take part, and study how the hyperons affect the properties of neutron stars including the maximum mass.  To perform such calculations, it is necessary to improve the conventional method for the DBHF calculation.  

First, we have examined if the space part of baryon self-energies can be neglected even in dense matter.  Then, we have found that, as we expect, such neglect is certainly valid around $n_B^0$, but that it grows rapidly above $2n_B^0$ and thus cannot be ignored at higher densities.  

Secondly, in the present calculation, we have partly included the contribution of baryon negative-energy states, because we want to remove the ambiguity in the relationship between the in-medium reaction matrices for baryon-baryon scattering and the baryon self-energies \cite{Fuchs1,Fuchs2,Fuchs3}.  It is meaningful to notice that, up to around $n_B^0$, the neglect of negative-energy states seems practically reasonable (although the ambiguity still remains), and that the result calculated by the subtraction scheme with the ps representation \cite{Fuchs2,DBHF3} is very similar to the present one.  

Furthermore, in the usual DBHF method, it is often assumed that the difference between the interacting nucleon masses can be ignored even in isospin-asymmetric nuclear matter \cite{DBHF3,Dalen}.  However, in general, the difference between those masses may not be negligible in the case with large isospin-asymmetry. 
In matter including hyperons, such mass difference becomes inherently large.  Thus, it is unavoidable to take into account the baryon-mass difference in dense medium \cite{Huber}. 

To manage these problems, in the present paper, we have solved the Bethe-Salpeter equations, Eqs.(\ref{eq_BS_for_T})-(\ref{eq_BS_for_P}), in the nuclear-matter rest frame by using the standard spinors, Eqs.(\ref{eq_positive_spinor})-(\ref{eq_negative_spinor}).  Note that, as in Ref.\cite{without_PW}, we do not adopt the partial-wave decomposition in the actual calculation.  Then, the baryon self-energies, the energy density and pressure of matter have been calculated using the Bonn potentials together with the hyperon-meson coupling constants determined in SU(6) symmetry.  

We have studied the EoS for neutron-star matter including hyperons in the following, three cases: (1) five baryons (neutron, proton, $\Lambda$, $\Sigma^-$, $\Xi^-$) are considered, and they interact with one another through the exchanges of six kinds of nonstrange mesons ($\sigma,\,\delta,\,\omega,\,\rho,\,\eta,\,\pi$), (2) eight baryons (neutron, proton, $\Lambda$, $\Sigma^-$, $\Sigma^0$, $\Sigma^+$, $\Xi^-$, $\Xi^0$) and the six nonstrange mesons are taken into account, (3) the five baryons take part in, but they interact through the exchanges of eight kinds of mesons ($\sigma,\,\delta,\,\omega,\,\rho,\,\eta,\,\pi,\,K,\,K^{\ast}$).  The last case includes the diagrams describing the exchange of interacting two baryons and the baryon transition.  For example, in the $N\Lambda$ channel, it contains the transition process, $N\Lambda\to N\Sigma$, as well as the diagonal one, $N\Lambda\to N\Lambda$.   

The particle fractions in neutron stars in the case (1) is very similar to that in the case (2), where only two hyperons ($\Lambda$, $\Xi^-$) appear in matter below $n_B=0.8\,\mathrm{fm}^{-3}$, and any $\Sigma$ hyperon does not emerge.  This tendency can be also seen in the result with RH or RHF approximation \cite{ApJ,miyatsu}.  The hindrance to the $\Sigma$ may be partly explained by the fact that the $\Sigma$-$\sigma$ coupling constant is chosen so as to reproduce the observed, repulsive potential at $n_B^0$.  

Therefore, in the last case (3), together with the full members of the mesons, we have considered only the five baryons. 
In the result of the case (3), we then have found that the inclusion of the $K$ and $K^\ast$ mesons leads to the appearance of the $\Sigma^-$ at $n_B\simeq0.38\,\mathrm{fm}^{-3}$ in neutron-star matter, which however dwindles rapidly at higher densities.  In contrast, the numbers of the $\Lambda$ and $\Xi^-$ grow steadily, once they emerge in matter.  

The calculation in the case (3) has predicted the maximum neutron-star mass of $2.02\,M_{\odot}$, which is consistent with both the recently observed masses, $1.97\pm0.04M_{\odot}$ (J1614-2230) and $2.01\pm0.04M_{\odot}$ (J0348+0432).  

Finally, we comment on the future work.
In our calculations, we have simply adopted the hyperon-meson coupling constants in SU(6) symmetry, and assumed that a cutoff mass in the hyperon form factor at the interaction vertex is taken to be the same value as in the nucleon form factor.
However, if those parameters are replaced with more realistic values \cite{Holzenkamp,Haidenbauer,Rijken}, the EoS for dense nuclear matter may be changed \cite{ApJ}.  
In Ref.\cite{Holzenkamp}, assuming that the hyperon-meson coupling constants (apart from those involving the fictitious $\sigma-$meson) are determined by SU(6) symmetry, the cutoff masses and the hyperon-$\sigma$ coupling constants have been studied using the experimental data of nucleon-hyperon scattering. 
Thus, it is very interesting to perform a calculation with more realistic coupling constants and cutoff masses in the present DBHF approach, and see how the EoS and the maximum neutron-star mass are modified. 

At very high densities, the quark and gluon degrees of freedom, rather than the hadron degrees of freedom, may take place in neutron-star matter \cite{Glendenning}.  Furthermore, in the crossover between the hadron and quark-gluon phases, there may exist rich non-perturbative structure such as color superconducting phases etc.
It would be very interesting to investigate, together with the DBHF calculation, how such degrees of freedom contribute to the EoS and the maximum mass of neutron star.

\vspace{1cm}
\begin{acknowledgements}
This work was supported by JSPS KAKENHI Grant Number 255742.
\end{acknowledgements}

%
%

%

\end{document}